\documentclass[twoside,11pt]{article}

\usepackage[arxiv-submission]{melba} 

\usepackage{mwe} 

\usepackage{amsmath,amsfonts}

\usepackage{subcaption}
\usepackage{makecell}

\usepackage{hyperref}
\usepackage{siunitx}

\newcommand{\etal}{\textit{et al}. }
\newcommand{\ie}{\textit{i}.\textit{e}., }
\newcommand{\eg}{\textit{e}.\textit{g}. }

\newcommand\sevenimagewidth{0.235\textwidth}

\melbaheading{2022:NNN}{https://www.melba-journal.org/article/XX-AA}{2022}{1-?}{m1/yyyy}{m2/yyyy}{G\"obl, Hennersperger and Navab}{}{}

\ShortHeadings{Speckle2Speckle: Unsupervised Learning of Ultrasound Speckle Filtering}{G\"obl, Hennersperger and Navab}
\firstpageno{1}

\title{Speckle2Speckle: Unsupervised Learning of Ultrasound Speckle Filtering Without Clean Data}

\author{\name R\"udiger G\"obl \email r.goebl@tum.de \\  
	\addr {Computer Aided Medical Procedures, Technical University Munich, Munich, Germany} \\
	\addr {OneProjects Design Innovation GmbH, Munich, Germany}
	\AND
	\name Christoph Hennersperger \email{christoph.hennersperger@tum.de} \\
	\addr {Computer Aided Medical Procedures, Technical University Munich, Munich, Germany} \\
	\addr {OneProjects Design Innovation GmbH, Munich, Germany}
	\AND
	\name Nassir Navab \email nassir.navab@tum.de \\
	\addr {Computer Aided Medical Procedures, Technical University Munich, Munich, Germany}
}

\begin{document}

\maketitle

\begin{abstract}
In ultrasound imaging the appearance of homogeneous regions of tissue is subject to speckle, which for certain applications can make the detection of tissue irregularities difficult. 
To cope with this, it is common practice to apply speckle reduction filters to the images. 
Most conventional filtering techniques are fairly hand-crafted and often need to be finely tuned to the present hardware, imaging scheme and application. 
Learning based techniques on the other hand suffer from the need for a target image for training (in case of fully supervised techniques) or require narrow, complex physics-based models of the speckle appearance that might not apply in all cases. 
With this work we propose a deep-learning based method for speckle removal without these limitations. 
To enable this, we make use of realistic ultrasound simulation techniques that allow for instantiation of several independent speckle realizations that represent the exact same tissue, thus allowing for the application of image reconstruction techniques that work with pairs of differently corrupted data.
Compared to two other state-of-the-art approaches (non-local means and the Optimized Bayesian non-local means filter) our method performs favorably in qualitative comparisons and quantitative evaluation, despite being trained on simulations alone, and is several orders of magnitude faster.
Our code is available at~\url{https://github.com/goeblr/Speckle2Speckle}.
\end{abstract}


\section{Introduction}
\label{sec:introduction}
\begin{figure}[t]
     \centering
     \begin{subfigure}[b]{\sevenimagewidth}
         \centering
         \includegraphics[width=\textwidth,trim=0 45 0 11, clip]{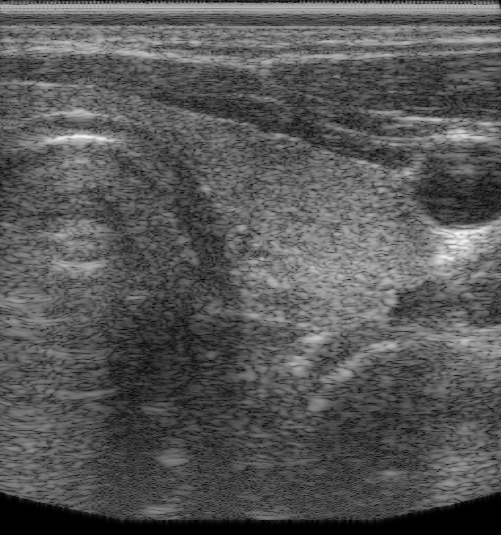}
         \caption{Unfiltered}
         \label{fig:thyroid-comparison-a}
     \end{subfigure}
     \hfill
     \begin{subfigure}[b]{\sevenimagewidth}
         \centering
         \includegraphics[width=\textwidth, trim=0 45 0 11, clip]{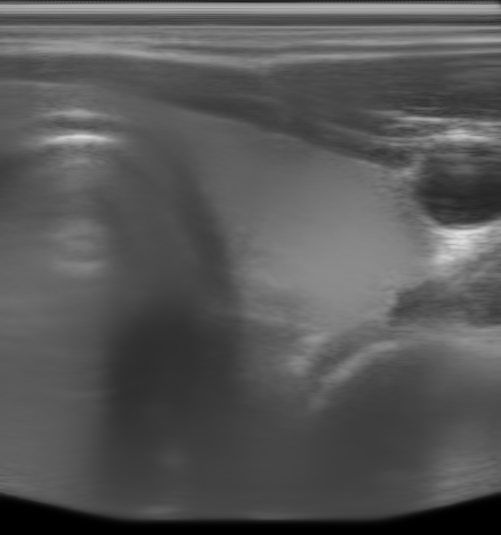}
         \caption{Filtered: OBNLM}
         \label{fig:thyroid-comparison-zoom-b}
     \end{subfigure}
     \hfill
     \begin{subfigure}[b]{\sevenimagewidth}
         \centering
         \includegraphics[width=\textwidth,trim=0 21 0 0, clip]{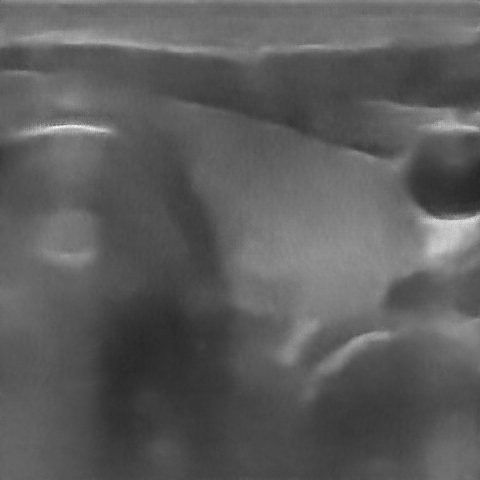}
         \caption{Filtered: Ours}
         \label{fig:thyroid-comparison-c}
     \end{subfigure}
     \hfill
     \begin{subfigure}[b]{\sevenimagewidth}
         \centering
         \includegraphics[width=\textwidth,trim=0 21 0 0, clip]{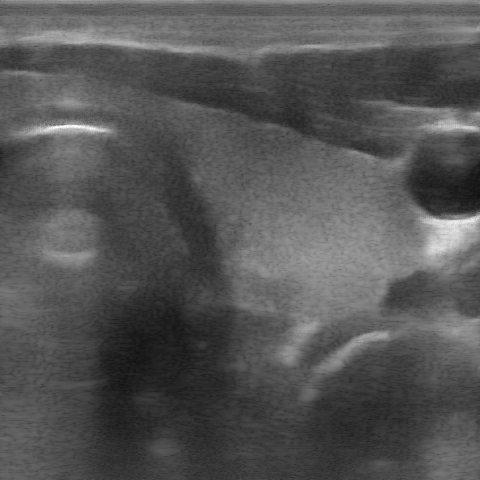}
         \caption{Blending}
         \label{fig:thyroid-comparison-d}
     \end{subfigure}
    \caption{Comparison of image appearance of a thyroid. 
    Unfiltered (a), despeckling with a state-of-the-art method (OBNLM) (b), despeckling with our method (c) and a linear combination of (a) and (c) in (d).}
    \label{fig:thyroid-comparison}
\end{figure}

Ultrasound imaging is a valuable tool to support non-invasive diagnosis and for use during interventions, as it is portable, affordable and does not subject the patient or staff to ionizing radiation.
It is not, however, without problems.
One of the major differences between CT and MRI on the one hand and ultrasound on the other hand is the presence of speckle, a grainy, noiselike image degradation that even affects tissue regions that are homogeneous in their microstructure such as liver or thyroid tissue (\eg see \autoref{fig:thyroid-comparison-a}).
This speckle can make interpretation of the images difficult---especially for users with limited experience working with ultrasound.
Consequently, most---if not all---ultrasound systems apply speckle reduction techniques after image formation, where the amount of "graininess" that is retained usually can be controlled by the user through a linear combination or blending of filtered and unfiltered data, depending on the clinical task and user preference.
\autoref{fig:thyroid-comparison} shows a thyroid image before and after filtering with our method compared to a state-of-the-art method and outlines a way to preserve a controlled amount of speckle.

The speckle patterns that emerge in ultrasound images are directly influenced by the location of sub-resolution scatterers in the medium.
Being (spaced) sub-resolution, the reflections caused by the individual scatterers cannot be separated by the ultrasound imaging system.
The result is interference between the echoes of different amplitude and phase, leading to an apparent echogenicity that changes depending on the spatial configuration of the scatterers within each resolution cell and the orientation of transmitting and receiving elements.
In addition to speckle, there is also electronic noise that degrades ultrasound images, which is not the focus of this work.
Speckle, while sometimes referred to as noise, is a deterministic phenomenon.
In fact, if it were possible to place an ultrasound transducer at the exact same location when repeating a scan, the resulting speckle pattern would be exactly the same.

\section{Related Works and Contribution}
Conventional despeckling approaches such as speckle-reducing anisotropic diffusion (SRAD) \citep{SRAD_yu2002speckle} and the Optimized Bayesian non-local means filter (OBNLM) \citep{OBNLM_coupe2009nonlocal}---itself building upon non-local means (NLM) \citep{NLM_buades2005non}---have been studied extensively.
We refer the interested reader to the review from \cite{DENOISING_REVIEW_sagheer2020review}.
Especially the latter method OBNLM exhibits excellent despeckling performance, but requires application-specific tuning and significant processing time.

\cite{ESAOTE_cammarasana2021universal} alleviate both these concerns by using a regression CNN that is trained against images despeckled with a tuned, computationally expensive conventional method, thereby allowing the CNN to replicate its results while lowering the execution time.

In other areas different to ultrasound imaging, there exists a body of work on the task of image denoising without requiring perfectly matched pairs of corrupted and clean data.
In Noise2Noise \citep{Lehtinen2018}, a CNN learns to restore images from a number of different corruptions, by using a corrupted image as input and a second image of the same "scene" with a different realization of the corruption as training target.
The authors show that this can reach the performance of methods training against clean images.
Noise2Void \citep{NOISE2VOID_krull2019noise2void} takes this even further by requiring only unpaired corrupted images, it requires however, strong assumptions on the corruption, namely zero-mean and per-pixel independent noise.

Given the difficulties of acquiring clean images in many medical imaging modalities, these general approaches have been further specialized.
In optical coherence imaging, a few well-chosen modifications to the acquisition setup can allow for the acquisition of two independent speckle instances with otherwise unchanged scanning setup \citep{OPTICAL_yin2020speckle}, effectively allowing for direct use of Noise2Noise.
However, this method cannot be directly applied to ultrasound imaging, as these modifications are not possible in ultrasound scanning setups.

In this work, we build on the concept of exploiting two independent \emph{corrupted} observations with the same scanning setup and geometry, and transfer this to a generalizable method for ultrasound imaging. 
In this way, we present a technique for learning a CNN-based ultrasound despeckle filter only from images containing speckle.
To do so, we apply the concept of Noise2Noise \citep{Lehtinen2018}, combined with a scheme for ultrasound simulation for data generation, and modifications to the training loss to enhance the image appearance in a method we call Speckle2Speckle.
By simulating different observations of the same general anatomical geometry but with different scatterer locations, we create images with uncorrelated speckle patterns.
In our qualitative comparison and quantitative evaluation with state-of-the-art approaches, we show that the method can remove speckle from real ultrasound data efficiently while not relying on manual data annotation or acquisitions, while the application of the filter is several orders of magnitude faster.

\section{Method}
\begin{figure}[htb]
    \setlength{\belowcaptionskip}{\baselineskip}
    \centering
    \begin{subfigure}[b]{\columnwidth}
        \centering
        \includegraphics[scale=0.55,page=1]{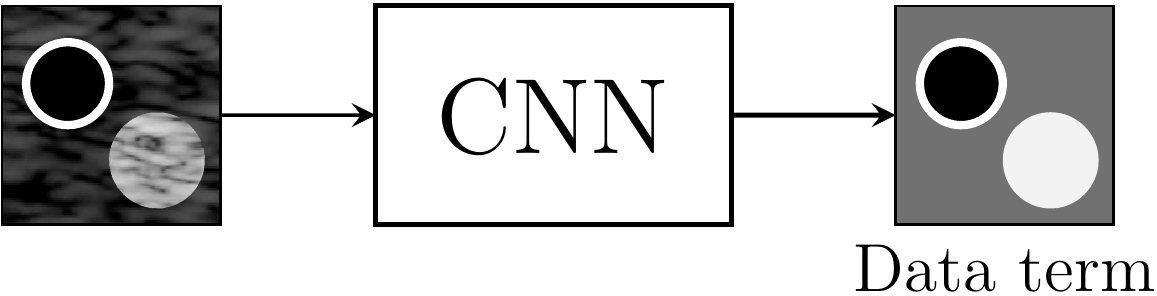}
        \caption{Clean target}
        \label{fig:method-overviews_a}
    \end{subfigure}
    \hfill
    \begin{subfigure}[b]{\columnwidth}
        \centering
        \includegraphics[scale=0.55,page=2]{images/Method_graphic-crop.pdf}
        \caption{Noise2Noise}
        \label{fig:method-overview_b}
    \end{subfigure}
    \par
    \begin{subfigure}[b]{\columnwidth}
        \centering
        \includegraphics[scale=0.55,page=3]{images/Method_graphic-crop.pdf}
        \caption{Ours}
        \label{fig:method-overview_c}
    \end{subfigure}
    \caption{Overview of training a model with corrupted data input to clean data output (a), Noise2Noise training with both corrupted data inputs and outputs of different realizations (b), and Speckle2Speckle training with corrupted realizations of ultrasound data simulated from the same tissue model (c).}
    \label{fig:method-overview}
\end{figure}

In this section we first summarize the original formulation of Noise2Noise, followed by a description how we create simulated ultrasound images to be used with a denoising approach like this, and the modifications that were required to account for the peculiarities of ultrasound.
\autoref{fig:method-overview} shows a comparison between the concepts for conventional (clean-target) \citep{ESAOTE_cammarasana2021universal}, Noise2Noise and Speckle2Speckle filter learning.
 
\subsection{Noise2Noise}
\label{subsec:noise2noise}
Noise2Noise \citep{Lehtinen2018} is a deep-learning based technique for general image restoration that does not require clean, uncorrupted data for training.
In a conventional restoration approach, one would use corrupted images as input and clean images of the same scene as the training target.
In their work however, Lehtinen \etal derive, from a statistical perspective, that a network trained with infinite samples can learn to estimate the expectation of the target samples.
This allows the input images as well as the target images to be corrupted with noise, like image noise observed with digital image sensors.
The authors show that the networks trained in this manner perform on-par with networks trained against clean data.

\begin{figure}[htb]
    \centering
    \includegraphics[width=0.9\textwidth]{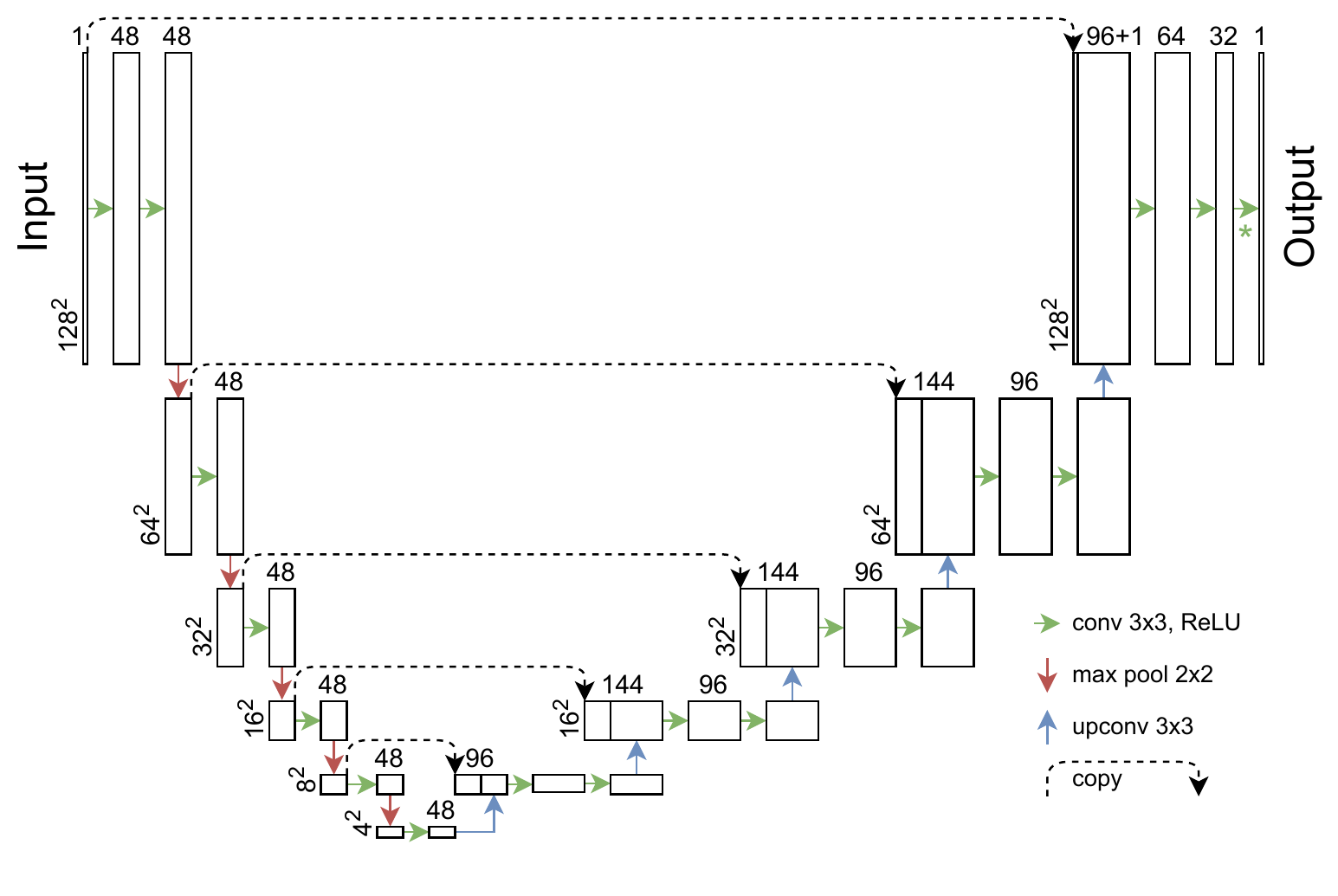}
    \caption{Network architecture of Noise2Noise and Speckle2Speckle. Numbers left of each box show the sizes of the feature maps, exemplified with a 128 by 128 pixel input. The number of channels per feature map is given by the numbers above the boxes. Different arrows represent different operations. The rightmost convolution (marked with $*$) has linear activation.}
    \label{fig:architecture}
\end{figure}

These relaxed requirements on the training targets allow for the application of this technique to situations where the acquisition of clean targets is either costly or impossible all-together.
In their work, Noise2Noise using a modified U-Net (following \cite{UNET_ronneberger2015u}, see \autoref{fig:architecture}) is applied to a number of different image corruption scenarios ranging from photographs subjected to Gaussian-, Poisson- and Bernoulli-noise or text-overlays, Monte-Carlo rendered images, and even MRI acquisitions within the randomly sampled k-space.

\subsection{Speckle2Speckle: Data}
\label{subsec:speckle2speckle_data}
Ultrasound imaging is a perfect example for a situation where a noise-free target image is difficult to obtain: since diffuse scattering is present for any ultrasound acquisition, speckle patterns remain visible at any time for acquired echo data.
Thus, one can argue that in ultrasound imaging, there is no image without speckle.
The strong correlation between the location and scattering behaviour of the tissue inhomogeneities causing the echos and the locations of extrema in the observed reflections (\ie the speckle pattern), however, makes the acquisition of multiple images with independent corruption (of the desired echogenicity measurement) difficult for real acquisitions.

\begin{figure}[htb]
     \centering
     \begin{subfigure}[b]{\sevenimagewidth}
         \centering
         \includegraphics[width=\textwidth]{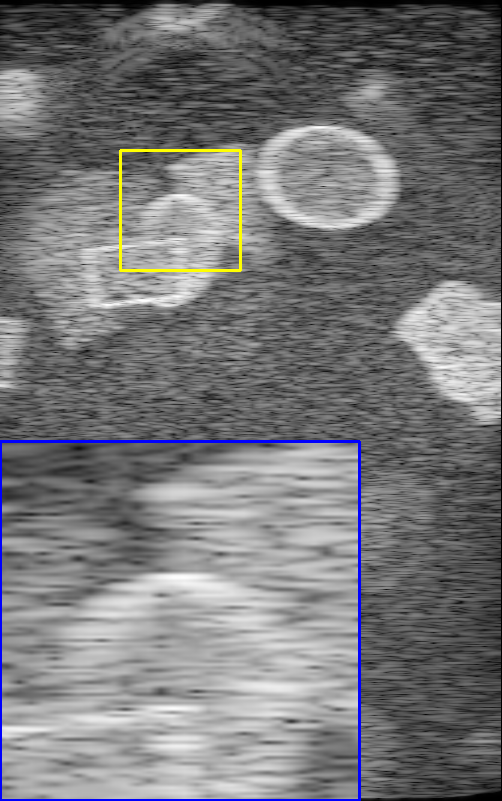}
         \caption{Instance 1}
         \label{fig:insilico_phantom_images_a}
     \end{subfigure}
     \hfill
     \begin{subfigure}[b]{\sevenimagewidth}
         \centering
         \includegraphics[width=\textwidth]{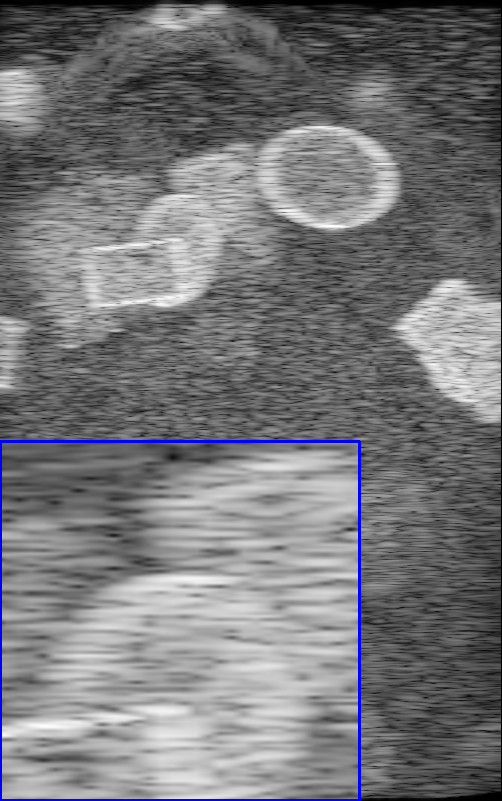}
         \caption{Instance 2}
         \label{fig:insilico_phantom_images_b}
     \end{subfigure}
     \hfill
     \begin{subfigure}[b]{\sevenimagewidth}
         \centering
         \includegraphics[width=\textwidth]{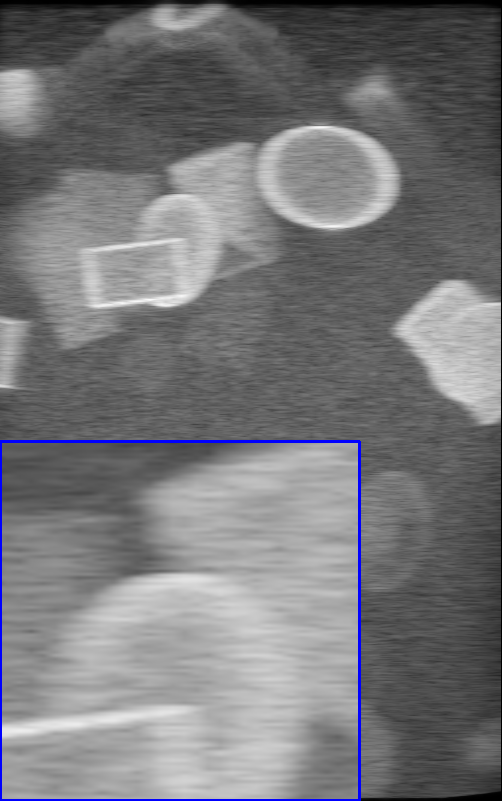}
         \caption{Average}
         \label{fig:insilico_phantom_images_mean}
     \end{subfigure}
     \hfill
     \begin{subfigure}[b]{\sevenimagewidth}
         \centering
         \includegraphics[width=\textwidth]{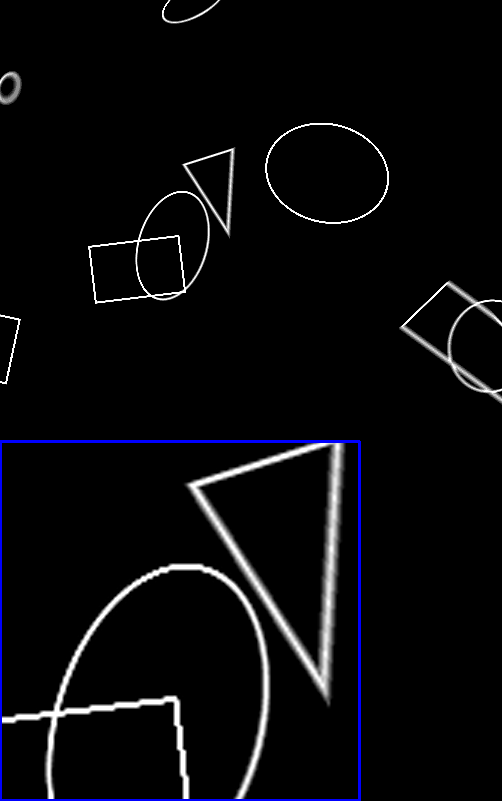}
         \caption{Interface map}
         \label{fig:insilico_phantom_images_interfaces}
     \end{subfigure}
        \caption{Comparison of the different image types created through simulation: 
        Ultrasound images of two different scatterer instances from the same phantom geometry (a) and (b), average of nine ultrasound images from the same geometry (c), interface map derived from the phantom geometry (d).}
        \label{fig:insilico_phantom_images}
\end{figure}

However, we can instead turn to realistic ultrasound simulations, where we can exploit the specific control over individual scatterers to generate an arbitrary amount of corrupted datasets for a given setup.
The training data for Speckle2Speckle is generated with the simulation software Field II \citep{jensen1992, jensen1996}.
Images are generated in pairs, using independent in-silico scatterer phantom datasets (collections of scatterers with associated scattering strengths).
These in-silico scatterer phantom datasets are derived from the same geometric phantom, which is a collection of 3D shapes each of which with a number of parameters describing their appearance in ultrasound imaging and scatterer parameters:
\begin{itemize}
    \item scatterer density
    \item scatterer amplitude distribution
    \item presence or absence of an interface
    \item interface scatterer density
    \item interface scatterer amplitude distribution
\end{itemize}
Based on the these properties, this approach allows for the generation of scatterer phantom datasets with independent scatterer locations and thus ideally suited for approaches such as Noise2Noise.
Scatterer locations can be generated randomly, but since the "macro-scale" geometry (e.g. areas of higher scattering) is the same, the ultrasound images created from the different scatterer phantoms look structurally identical and even the speckle distributions are identical.
The main difference, however, is that the speckle pattern "instance" is different, for which an example is shown in \autoref{fig:insilico_phantom_images_a} and \autoref{fig:insilico_phantom_images_b}.

While the training is performed with pairs of scatterer instances, we use a different target for the testing and validation.
In order to facilitate the evaluation of speckle removal in the absence of noise-free images, we compare the output of our method with an image that is composed by averaging multiple ultrasound images of the same phantom geometry but different scatterer instances.
For the training and validation sets, we generated 10 instances per phantom, one of which is used (randomly) as input, the other nine are averaged.
\autoref{fig:insilico_phantom_images_mean} shows the average of nine images corresponding to the same geometry as in \autoref{fig:insilico_phantom_images_a} and \autoref{fig:insilico_phantom_images_b}.

In addition to the ultrasound images, we also derive an interface map directly from the geometric phantom.
Being of the same size and spatial origin as the simulated ultrasound images, it contains whether an interface of a shape was present at each location.
\autoref{fig:insilico_phantom_images_interfaces} shows the interface map associated with the geometry used for the simulations of the ultrasound images in \autoref{fig:insilico_phantom_images}.
This is later utilized to adapt the training loss to the challenges of ultrasound imaging in order to achieve enhancements of prominent interfaces in the output.

Following this general scheme, a dataset was generated with a training set consisting of 1000 image pairs while the validation and test sets each contain 100 phantom geometries with 10 speckle instances per phantom geometry, resulting in 4000 simulated images in total.
Fundamentally, the size of the dataset was limited by the simulation time, where each image required roughly between one and two days on a single CPU core to simulate (depending on the concrete number of scatterers and the CPU performance), although parallelization between images was trivial.

In order to limit the simulation run-time, the scatterer densities values were determined as high as required such that further increase did not cause a change in the speckle patterns.
The probabilities of geometries being anechoic and the scatterer amplitude distributions were determined arbitrarily to ensure visibility of the different regions.
The specific parametrization is listed in \autoref{tab:phantomparameters}.
\begin{table}[t]
\centering
\begin{tabular}{l|l|l}
\hline
Region              & Parameter                      & Value                                                                                                                             \\ \hline
Background          & probability anechoic           & 0.4                                                                                                                               \\
                    & scatterer density              & \SI{0.333e10}{\meter^{-3}}                                                                                                              \\
                    & scatterer amplitude            & \makecell[tl]{normal distributed $\mathcal{N}(0, \sigma_{\text{bg}}^2)$; \\ $\sigma_{\text{bg}}$ drawn per phantom from $\mathcal{N}(1, 0.5^2)$}    \\ \hline
Inclusion           & random objects                 & spheroids \& cuboids, uniformly distributed                                                                                       \\
                    & number per phantom             & 100                                                                                                                               \\
                    & extents                        & drawn per axis uniformly from \SI[number-math-rm = \mathnormal, parse-numbers = false]{[1, 5]}{mm}                                                                            \\ 
                    & probability anechoic           & 0.4                                                                                                                               \\
                    & scatterer density              & \SI{0.333e10}{\meter^{-3}}                                                                                                              \\
                    & scatterer amplitude            & \makecell[tl]{normal distributed $\mathcal{N}(0, \sigma_{\text{inc}}^2)$; \\ $\sigma_{\text{inc}}$ drawn per inclusion from $\mathcal{N}(4, 2^2)$} \\ \hline
Inclusion Interface & probability interface          & 0.5                                                                                                                               \\
                    & scatterer density              & \SI{5e6}{\meter^{-2}}                                                                                                                  \\
                    & scatterer amplitude            & \makecell[tl]{constant per inclusion,\\ drawn per inclusion from $\mathcal{N}(14, 2^2)$}                                       \\ \hline
\end{tabular}
\caption{Simulated phantom parameters and values}
\label{tab:phantomparameters}
\end{table}

The simulated imaging protocol used focused imaging and it followed the same restrictions as an actual focused scan using a CPLA12875 probe with a cQuest Cicada (both Cephasonics, CA, USA) would be subject to.
\autoref{tab:acquisitionparameters} contains the detailed acquisition parameters.
\begin{table}[htb]
\centering
\begin{tabular}{l|l}
\hline
Parameter                  & Value           \\ \hline
transducer                 & CPLA12875       \\
number elements            & 128             \\
element pitch              & \SI{0.30}{mm}   \\
elevation focus            & None            \\
transmit frequency         & \SI{7}{MHz}     \\
transmit pulse             & bipolar         \\
imaging depth              & \SI{60}{mm}     \\
focus depth                & \SI{30}{mm}     \\
steering                   & \SI{0}{\degree} \\
number scanlines           & 128             \\
TX aperture size           & 32              \\
RX aperture size           & 64              \\ \hline
\end{tabular}
\caption{Simulated acquisition parameters and values}
\label{tab:acquisitionparameters}
\end{table}

The output of the simulations was ultrasound radio-frequency channel data, which was subsequently processed to ultrasound images using SUPRA \citep{Goebl2018supra} with parameters as in \autoref{tab:reconstructionparameters}.
After beamforming, the resulting images were converted to PNGs and interface indicator maps for inclusions with an interface were created by examining the geometry at every pixel location for the presence of an interface.
\begin{table}[htb]
\centering
\begin{tabular}{l|l}
\hline
Parameter        & Value                    \\ \hline
beamforming      & DAS with hamming window  \\
dynamic range    & 70 dB                    \\
image resolution & 0.075 mm                 \\
image width      & \SI{37.6}{mm} $\widehat{=}$ 502 pixels \\
image height     & \SI{60.0}{mm} $\widehat{=}$ 801 pixels \\ \hline
\end{tabular}
\caption{Image reconstruction parameters and values}
\label{tab:reconstructionparameters}
\end{table}

\subsection{Speckle2Speckle: Learning}
The network architecture employed in Speckle2Speckle follows Noise2Noise (see \autoref{subsec:noise2noise}).

The training loss, however, was adjusted for the peculiarities of ultrasound imaging.
While the original Noise2Noise approach used different loss data terms depending on the noise, we extended the data term $L_2$ with a term to enhance the appearance of interfaces in the images, as interfaces retain key information for interpreting ultrasound imaging data.

The interface term is based on the known locations of interfaces given by the interface indicator map $I_\text{i}$.
We chose $I_{\text{i}G} = g_\text{i}*I_\text{i}$ as spatial weight for the interface loss, where $g_\text{i}$ is a 2D Gaussian with standard deviation $\sigma_i$.
This serves to i) extend the range of the interface effects from the potentially sharp peaks in the interface map, as well as ii) achieve a smooth transition between the effects of the data term and the interface term.
$\tilde{I}_{\text{i}G} = 1 - I_{\text{i}G}$ is the inverse weight.
With that, we formulate the overall loss $L_D$ as
\begin{align}
    &L_D (I_\text{o}, I_\text{t}, I_\text{i}) = 
    L_2 (I_\text{o} \odot \tilde{I}_{\text{i}G}, I_\text{t} \odot \tilde{I}_{\text{i}G}) + \lambda S_D (I_\text{o}, I_\text{t}, I_\text{i}),\\
    &S_D (I_\text{o}, I_\text{t}, I_\text{i} )= \frac{1}{|\Omega|} \sum_{x\in \Omega} \left((I_\text{o}*g_\text{psf} - I_\text{t}) \odot I_{\text{i}G}\right)^2, 
\end{align}
where $I_\text{o}$ is the network output, $I_\text{t}$ the target image, $\lambda$ the weighting of the interface term, $\odot$ denotes the Hadamard product (element-wise multiplication) and $g_\text{psf}$ a filter kernel of the point spread function (PSF). 
The notion is: In the target image, the interfaces are smoothed by the PSF by the process of imaging. 
Whereas in the output image the goal is to reduce the blurring, \ie increase the sharpness.
Consequently, the term $I_\text{o}*g_\text{psf} - I_\text{t}$ penalizes deviations in $I_\text{o}$ from sharp peaks.
The PSF in this case could be the result of a per-image estimation, learned for one specific system, manually determined for one system or treated as hyper-parameter. 
We assumed the PSF to be an anisotropic Gaussian, with $\sigma_\text{psf}^x = \text{FWHM}_\text{lat}/2$ and $\sigma_\text{psf}^y = \text{FWHM}_\text{ax}/2$ determined experimentally from a simulated point reflector, where we measured the full width half maximum (FWHM) in lateral and axial directions.
The hyper-parameters and their values are listed in \autoref{tab:hyperparameters}.
\begin{table}[htb]
\centering
\begin{tabular}{l|l}
\hline
Parameter                       & Value      \\ \hline
$\lambda$                       & \num{500}  \\
learning rate                   & \num{3e-5} \\
$\text{epochs}$                 & \num{1000} \\
$\sigma_\text{i}$ (interface)   & \SI{0.375}{mm} $\widehat{=}$ 5 pixels \\
$\sigma_\text{psf}^y$ (axial)   & \SI{0.075}{mm} $\widehat{=}$ 1 pixel \\
$\sigma_\text{psf}^x$ (lateral) & \SI{0.525}{mm} $\widehat{=}$ 7 pixels\\ \hline
\end{tabular}
\caption{Training hyper-parameters and their values}
\label{tab:hyperparameters}
\end{table}


\section{Evaluation}
In order to achieve a fair comparison we evaluate our approach to a group of well-established and state-of-the-art speckle filters, and provide qualitative comparisons and quantitative evaluations for the simulated, phantom, and in-vivo datasets in the following.
To this end, we compare the results of our method against SRAD\footnote{implementation from \cite{SRAD_code}} \citep{SRAD_yu2002speckle}, 
median filter\footnotemark{}, bilateral filter\footnotemark[\value{footnote}] \citep{BILAT_tomasi1998bilateral}, 
non-local means\footnotemark[\value{footnote}] (NLM) \citep{NLM_buades2005non}, \footnotetext{implementation from MATLAB R2020b}
and optimized Bayesian non-local means\footnote{implementation from \cite{OBNLM_code}} (OBNLM) \citep{OBNLM_coupe2009nonlocal}.
\autoref{tab:othermethod_parameters} contains the parametrization used for the mentioned methods.
\begin{table}[htb]
\centering
\begin{tabular}{l|l|l}
\hline
Method    & Parameter              & Value \\ \hline
SRAD      & number of iterations   & 200   \\
          & lambda                 & 0.1   \\ \hline
Median    & window size            & 15    \\ \hline
Bilateral & degree of smoothing    & 0.05  \\
          & spatial sigma          & 5     \\ \hline
NLM       & degree of smoothing    & 0.075 \\
          & search windows size    & 101   \\
          & comparison window size & 21    \\ \hline
OBNLM     & search area size       & 101   \\
          & patch size             & 45    \\
          & degree of smoothing    & 1.05  \\ \hline
\end{tabular}
\caption{Comparison method parameters and their values after manual tuning on the image in \autoref{fig:qualitative_comparison_invivo-a}.}
\label{tab:othermethod_parameters}
\end{table}

\subsection{Qualitative comparison}
\begin{figure}[htb]
\setlength{\belowcaptionskip}{\baselineskip}
     \centering
     \begin{subfigure}[b]{0.24\linewidth}
         \centering
         \includegraphics[width=\textwidth]{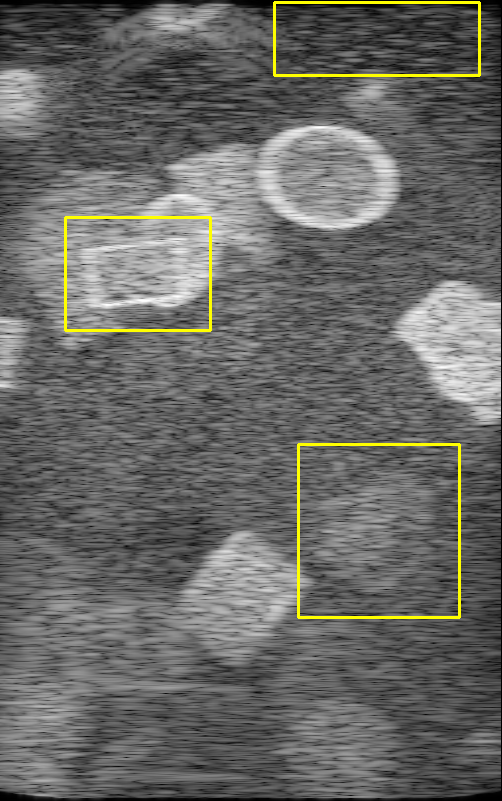}
         \caption{Original}
         \label{fig:qualitative_comparison_simulation-a}
     \end{subfigure}
     \hfill
     \begin{subfigure}[b]{0.24\linewidth}
         \centering
         \includegraphics[width=\textwidth]{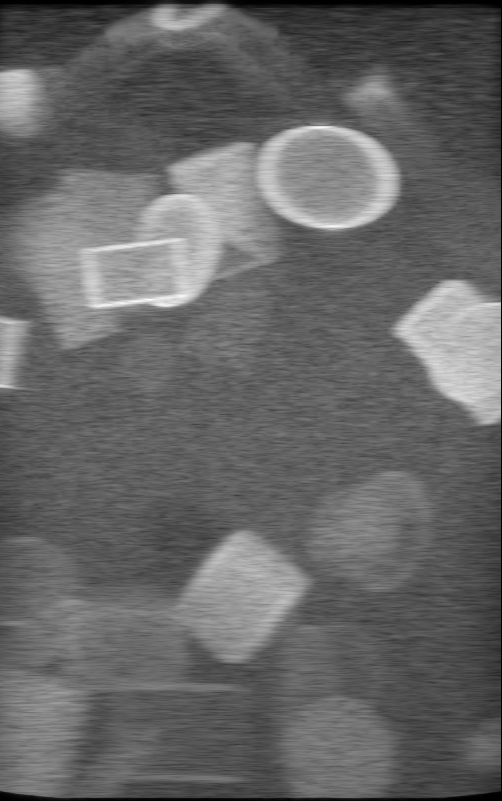}
         \caption{Average image}
         \label{fig:qualitative_comparison_simulation-b}
     \end{subfigure}
     \hfill
     \begin{subfigure}[b]{0.24\linewidth}
         \centering
         \includegraphics[width=\textwidth]{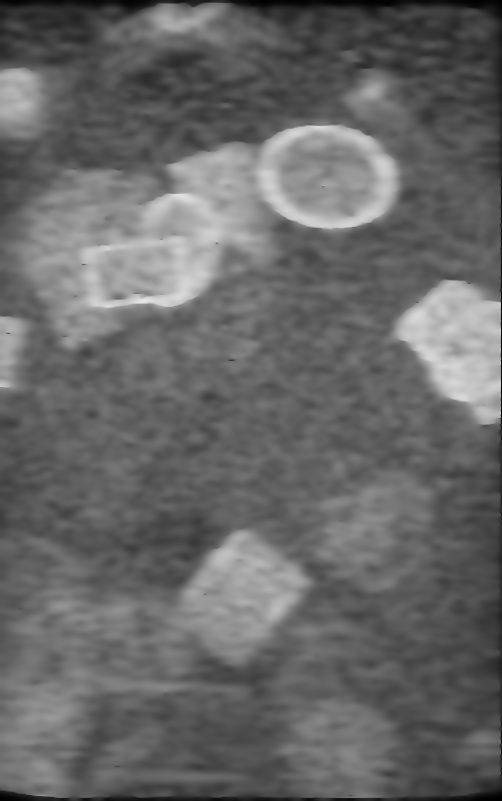}
         \caption{SRAD}
         \label{fig:qualitative_comparison_simulation-c}
     \end{subfigure}
     \hfill
     \begin{subfigure}[b]{0.24\linewidth}
         \centering
         \includegraphics[width=\textwidth]{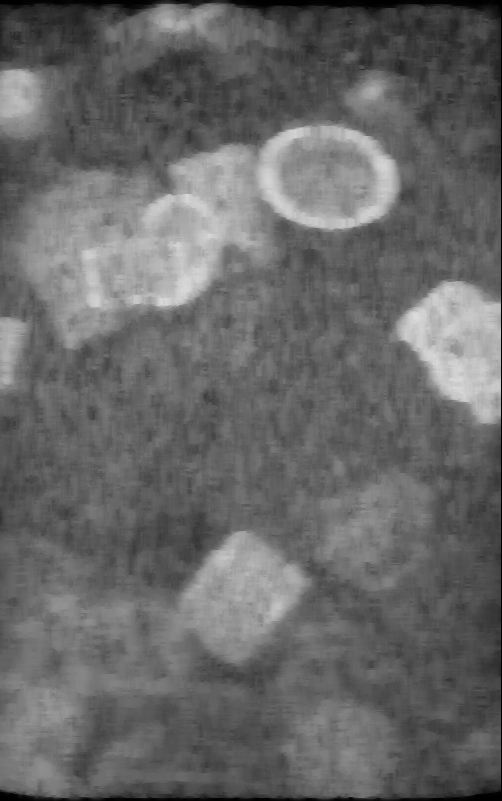}
         \caption{Median}
         \label{fig:qualitative_comparison_simulation-d}
     \end{subfigure}
     \par
     \begin{subfigure}[b]{0.24\linewidth}
         \centering
         \includegraphics[width=\textwidth]{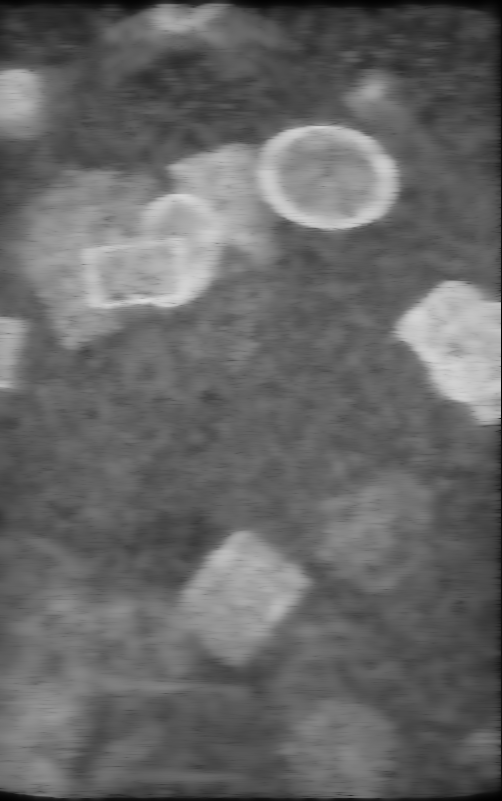}
         \caption{Bilateral}
         \label{fig:qualitative_comparison_simulation-e}
     \end{subfigure}
     \hfill
     \begin{subfigure}[b]{0.24\linewidth}
         \centering
         \includegraphics[width=\textwidth]{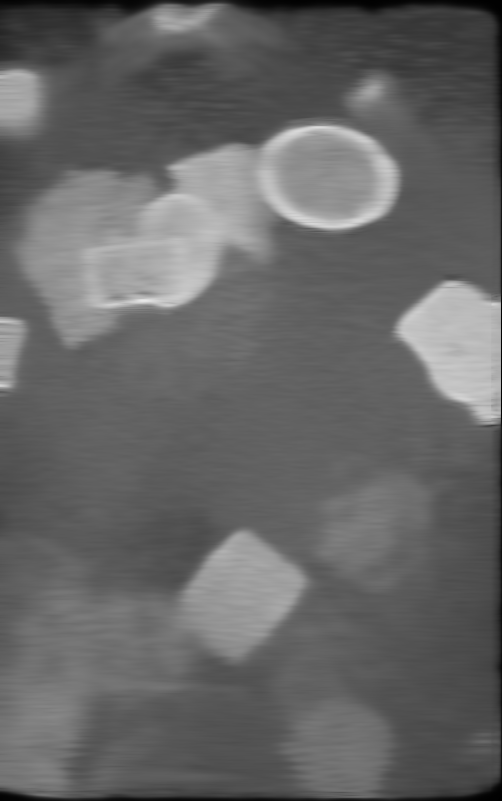}
         \caption{NLM}
         \label{fig:qualitative_comparison_simulation-f}
     \end{subfigure}
     \hfill
     \begin{subfigure}[b]{0.24\linewidth}
         \centering
         \includegraphics[width=\textwidth]{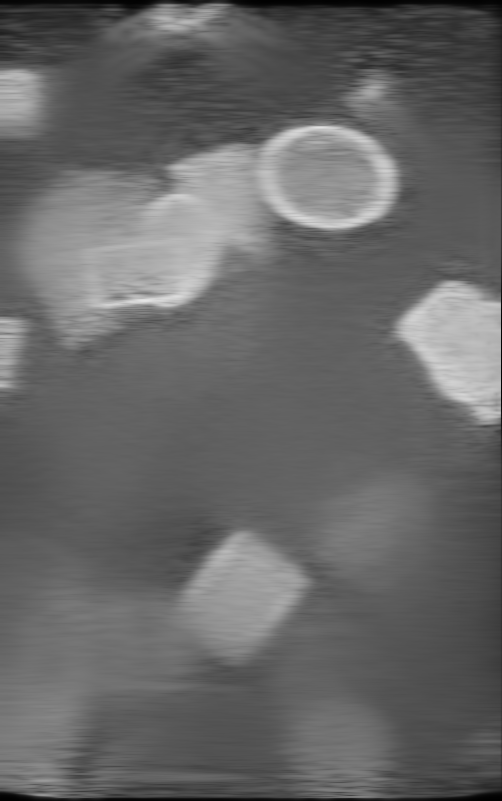}
         \caption{OBNLM}
         \label{fig:qualitative_comparison_simulation-g}
     \end{subfigure}
     \hfill
     \begin{subfigure}[b]{0.24\linewidth}
         \centering
         \includegraphics[width=\textwidth]{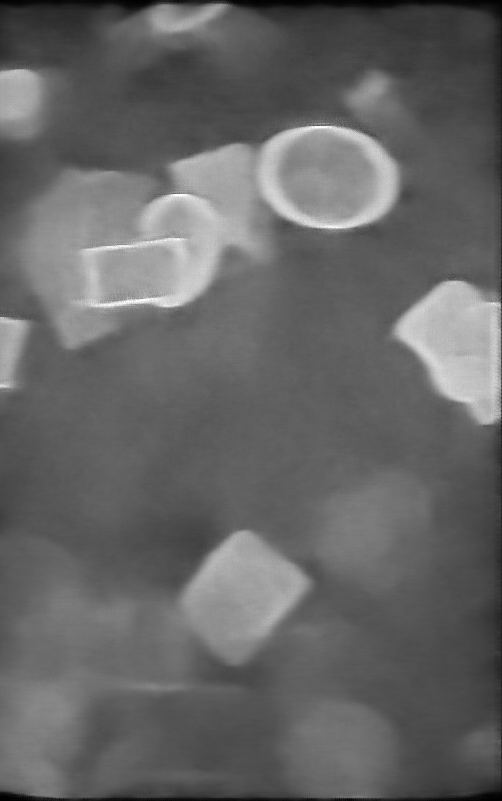}
         \caption{Ours}
         \label{fig:qualitative_comparison_simulation-h}
     \end{subfigure}
        \caption{
    Qualitative comparison on a simulated image. 
    The original speckled image (a), the average of nine different scatterer instantiations (b), output of SRAD (c), median filter (d), bilateral filtering (e), NLM (f), OBNLM (g) and ours (h).}
        \label{fig:qualitative_comparison_simulation}
\end{figure}
\emph{Simulated data.}
In the simulated image from the validation set (see \autoref{subsec:speckle2speckle_data}) shown in \autoref{fig:qualitative_comparison_simulation}, we can visually compare the result of applying the different methods directly with the average image of nine speckle instances.
It is clear that SRAD (c), median filtering (d) and the bilateral filter (e) mostly change the appearance of the speckle, but don not fully remove it.
NLM (f), OBNLM (g) and our method (h) all remove the speckle very well.
In the outlined area in the top of the image, the less dense speckle is not fully removed by OBNLM.
Another noteworthy region is the rectangular object in the top left that has a bright interface in the input as well as the average image.
NLM and OBNLM do not retain this feature with high contrast, whereas our method shows contrast close to the average image.
The elliptic object in the bottom right appears in our method not as clearly as it is visible in the average image, we believe this is due to comparatively low brightness of interface around it.
The other methods, however, also struggle to highlight the object.

\begin{figure}[htb]
\setlength{\belowcaptionskip}{\baselineskip}
     \centering
     \begin{subfigure}[b]{\sevenimagewidth}
         \centering
         \includegraphics[width=\textwidth]{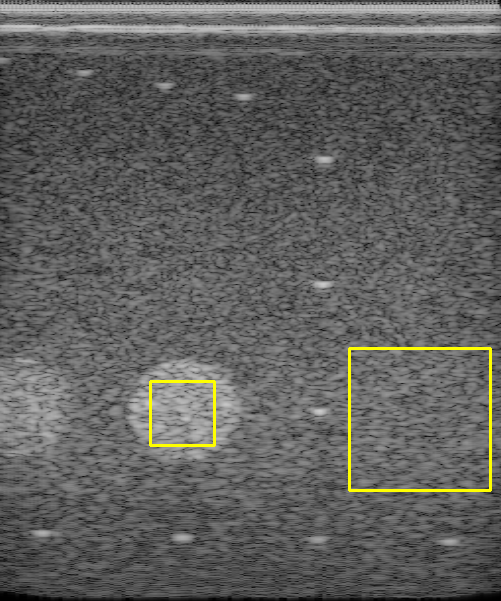}
         \caption{Original}
         \label{fig:qualitative_comparison_experimental-a}
     \end{subfigure}
     \hfill
     \begin{subfigure}[b]{\sevenimagewidth}
         \centering
         \includegraphics[width=\textwidth]{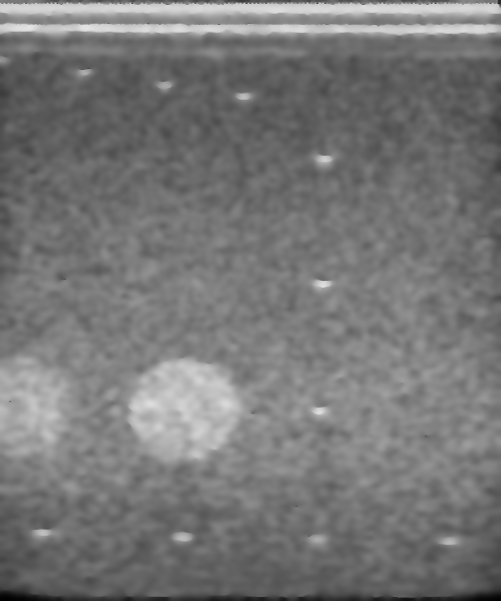}
         \caption{SRAD}
         \label{fig:qualitative_comparison_experimental-c}
     \end{subfigure}
     \hfill
     \begin{subfigure}[b]{\sevenimagewidth}
         \centering
         \includegraphics[width=\textwidth]{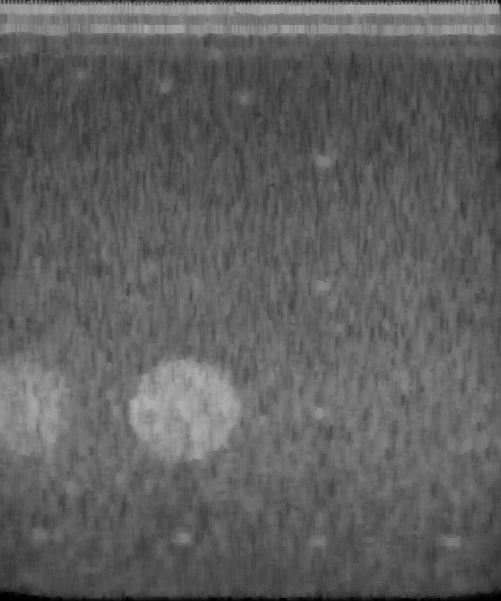}
         \caption{Median}
         \label{fig:qualitative_comparison_experimental-d}
     \end{subfigure}
     \hfill
     \begin{subfigure}[b]{\sevenimagewidth}
         \centering
         \includegraphics[width=\textwidth]{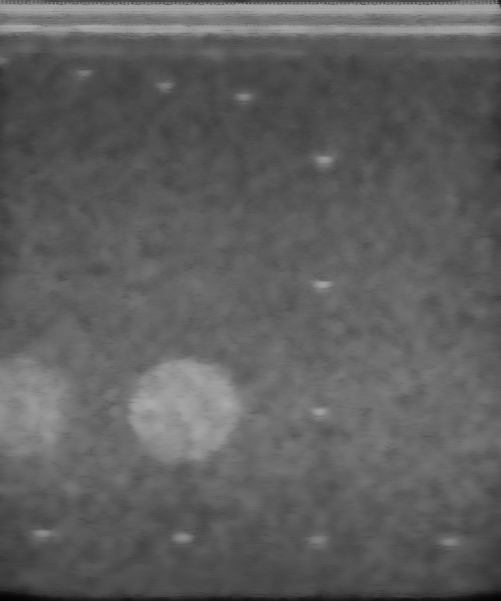}
         \caption{Bilateral}
         \label{fig:qualitative_comparison_experimental-e}
     \end{subfigure}
     \par
     \begin{subfigure}[b]{\sevenimagewidth}
         \centering
         \includegraphics[width=\textwidth]{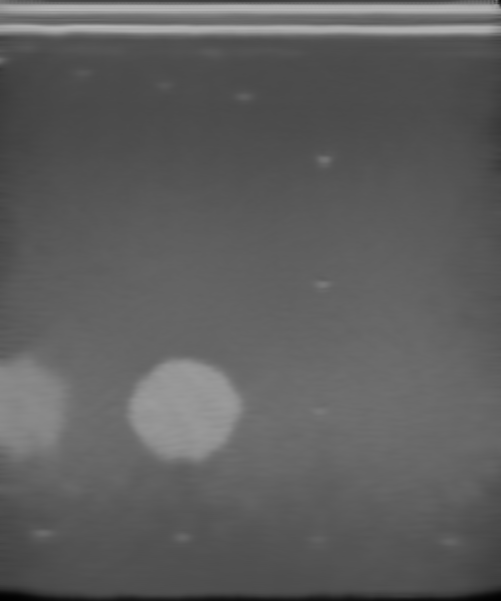}
         \caption{NLM}
         \label{fig:qualitative_comparison_experimental-f}
     \end{subfigure}
     \hfill
     \begin{subfigure}[b]{\sevenimagewidth}
         \centering
         \includegraphics[width=\textwidth]{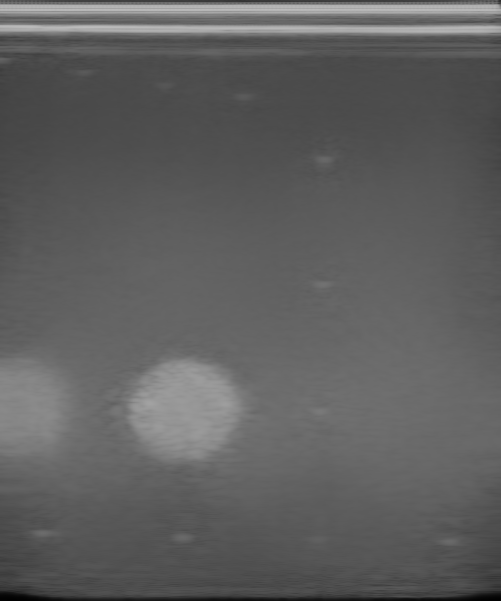}
         \caption{OBNLM}
         \label{fig:qualitative_comparison_experimental-g}
     \end{subfigure}
     \hfill
     \begin{subfigure}[b]{\sevenimagewidth}
         \centering
         \includegraphics[width=\textwidth]{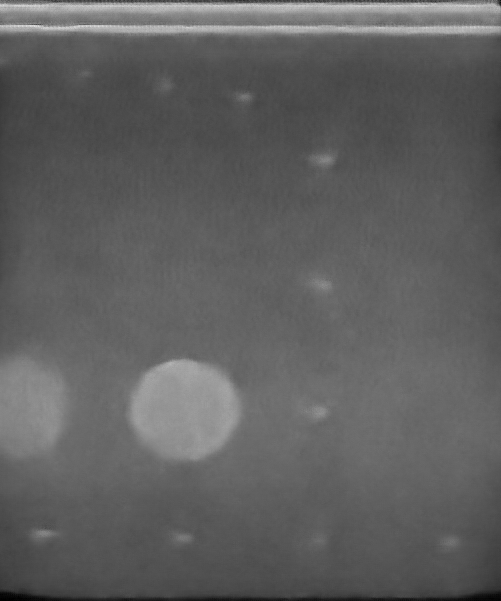}
         \caption{Ours}
         \label{fig:qualitative_comparison_experimental-h}
     \end{subfigure}
     \hfill
     \hspace{\sevenimagewidth}
        \caption{
    Qualitative comparison on an experimental phantom image. 
    The original speckled image (a), the output of SRAD (b), median filter (c), bilateral filtering (d), NLM (e), OBNLM (f) and ours (g). 
    The marked regions in the inclusion and background are used for quantitative evaluation.}
        \label{fig:qualitative_comparison_experimental}
\end{figure}
\emph{Phantom data.}
The experimental image in \autoref{fig:qualitative_comparison_experimental} shows an ultrasound quality assurance phantom (Model 040GSE, CIRS Inc., Norfolk, VA,
USA).
The image was acquired with a CPLA12875 probe and a cQuest Cicada (both Cephasonics, CA, USA), with equivalent imaging parameters as used for the simulations (see \autoref{tab:acquisitionparameters}), but with an imaging depth of \SI{45}{mm}.
Being acquired from a physical phantom, there is no average image as reference, but the distinct regions in the phantom are constructed as regions of homogeneous scattering.
The image shows a background region, several string reflectors, as well as two different circular hyper-echoic inclusions.
The result of SRAD (b), median (c) and bilateral (d) show the same general patterns as with the simulated data---not succeeding in completely removing the speckle.
NLM (e), OBNLM (f) and our method (g) are again close to that goal.
While NLM shows the smoothest image among the three methods and shows a sharp edge of the bright inclusion, it suppresses the string responses more.
The result of OBNLM is almost as smooth as NLM overall, but it shows inhomogeneities around the bright inclusion and its outline is not as well defined as with the other methods.
Our method---while smooth locally---shows a certain inhomogeneity in the background, but does not suppress the point reflectors as strongly and has a clear demarcation of both inclusions.

\begin{figure}[htb]
\setlength{\belowcaptionskip}{\baselineskip}
     \centering
     \begin{subfigure}[b]{\sevenimagewidth}
         \centering
         \includegraphics[width=\textwidth,trim=0 37 0 0, clip]{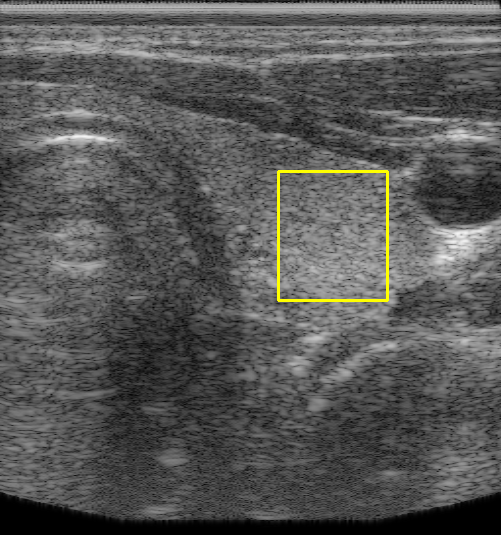}
         \caption{Original}
         \label{fig:qualitative_comparison_invivo-a}
     \end{subfigure}
     \hfill
     \begin{subfigure}[b]{\sevenimagewidth}
         \centering
         \includegraphics[width=\textwidth,trim=0 37 0 0, clip]{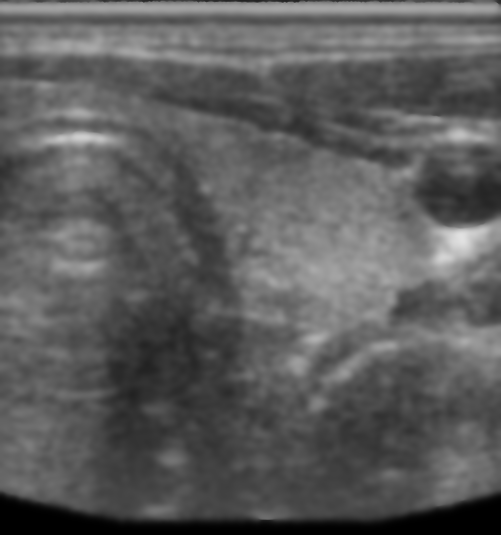}
         \caption{SRAD}
         \label{fig:qualitative_comparison_invivo-c}
     \end{subfigure}
     \hfill
     \begin{subfigure}[b]{\sevenimagewidth}
         \centering
         \includegraphics[width=\textwidth,trim=0 37 0 0, clip]{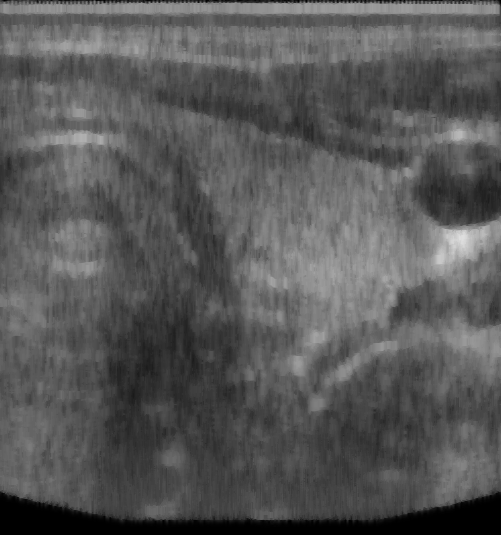}
         \caption{Median}
         \label{fig:qualitative_comparison_invivo-d}
     \end{subfigure}
     \hfill
     \begin{subfigure}[b]{\sevenimagewidth}
         \centering
         \includegraphics[width=\textwidth,trim=0 37 0 0, clip]{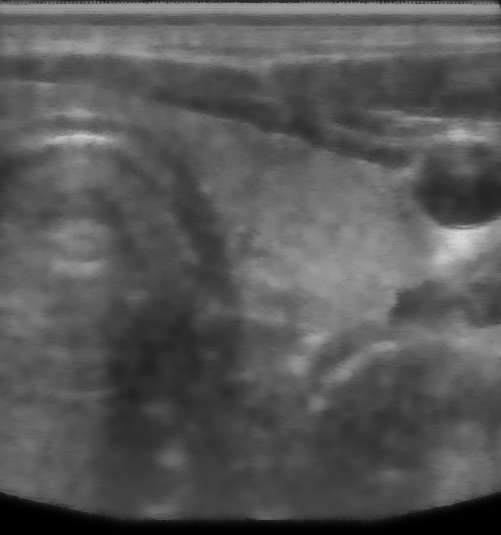}
         \caption{Bilateral}
         \label{fig:qualitative_comparison_invivo-e}
     \end{subfigure}
     \par
     \begin{subfigure}[b]{\sevenimagewidth}
         \centering
         \includegraphics[width=\textwidth,trim=0 37 0 0, clip]{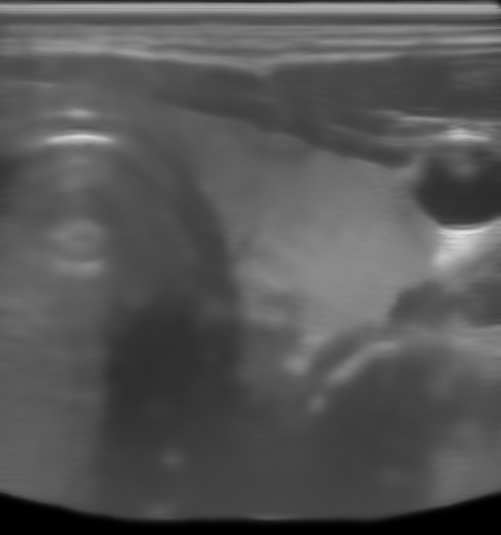}
         \caption{NLM}
         \label{fig:qualitative_comparison_invivo-f}
     \end{subfigure}
     \hfill
     \begin{subfigure}[b]{\sevenimagewidth}
         \centering
         \includegraphics[width=\textwidth,trim=0 37 0 0, clip]{images/qualitative_comparison_invivo/invivo_2021-01-17_image_16_1_OBNLM.png}
         \caption{OBNLM}
         \label{fig:qualitative_comparison_invivo-g}
     \end{subfigure}
     \hfill
     \begin{subfigure}[b]{\sevenimagewidth}
         \centering
         \includegraphics[width=\textwidth,trim=0 37 0 0, clip]{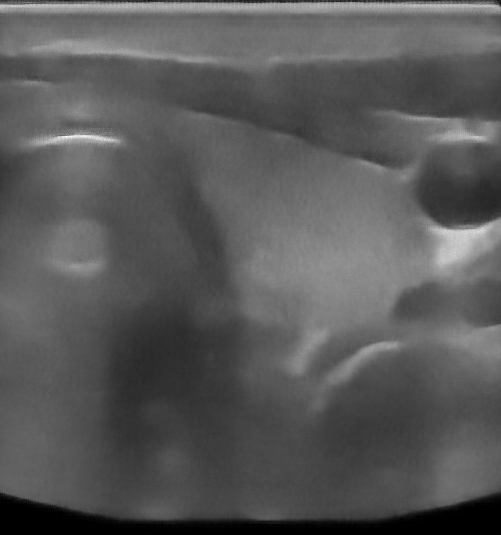}
         \caption{Ours}
         \label{fig:qualitative_comparison_invivo-h}
     \end{subfigure}
     \hfill
     \hspace{\sevenimagewidth}
        \caption{
    Qualitative comparison on an in-vivo image. 
    The original speckled image (a), the output of SRAD (b), median filter (c), bilateral filtering (d), NLM (e), OBNLM (f) and ours (g). 
    The marked region in the thyroid is used for quantitative evaluation.}
        \label{fig:qualitative_comparison_invivo}
\end{figure}
\emph{In-vivo data.}
In the in-vivo image in \autoref{fig:qualitative_comparison_invivo}---a cross-sectional view of the thyroid of a healthy volunteer acquired with the same configuration as the phantom data with \SI{35}{mm} depth---we can see that SRAD (b), median (c) and bilateral filtering (d) perform better than in the simulated and experimental cases, although NLM (e), OBNLM (f) and our method (g) exhibit a significantly more homogeneous appearance in the thyroid and especially OBNLM and our method retain more sharpness in comparison.
While OBNLM better captures the muscle interfaces over the carotid artery on the right of the image compared to our method, it is also less smooth in the overall area of the carotid.
Our method does not retain said interfaces completely, but is not disturbed by the clutter within the artery.

\subsection{Quantitative evaluation}
\label{subsec:quantitative_comparison}
\emph{Simulated data.}
For the quantitative evaluation in the simulated case we utilize the average images as an approximation of the speckle-free image.
We perform this for all 100 simulated images in the validation set and the corresponding average images.
The input images are scaled to the range $[0, 1]$ before application of any of the methods.
\autoref{tab:quantitative_comparison_simulation} shows the mean and standard deviation of the mean squared errors (MSE) and the mean absolute differences (MAD) between the considered methods and the average images.
\begin{table}[htb]
\centering
\begin{tabular}{l|r@{ }l|r@{ }l}
\hline
Method    & \multicolumn{2}{l|}{MSE ($\times 10^{-3}$)}        & \multicolumn{2}{l}{MAD ($\times 10^{-2}$)}  \\ \hline
Input     & 5.87           & $\pm$ 1.31           & 5.86           & $\pm$ 0.87           \\
SRAD      & 2.33           & $\pm$ 0.86           & 3.61           & $\pm$ 0.76           \\
Median    & 1.82           & $\pm$ 0.25           & 3.25           & $\pm$ 0.33           \\
Bilateral & 1.48           & $\pm$ 0.20           & 2.94           & $\pm$ 0.30           \\
NLM       & 1.27           & $\pm$ \bfseries 0.09 & 2.70           & $\pm$ \bfseries 0.13 \\
OBNLM     & 1.48           & $\pm$ 0.15           & 2.87           & $\pm$ 0.20           \\
Ours      & \bfseries 1.20 & $\pm$ 0.11           & \bfseries 2.60 & $\pm$ 0.20           \\ \hline
\end{tabular}
\caption{Quantitative comparison on 100 simulated images from the validation set. It shows the mean and standard deviation of the MSE and the MAD between images processed with each methods (and the unprocessed input) against the average images of nine speckle instances.
Lowest error mean and standard deviation for both MSE and MAD are highlighted in bold without implying significance.}
\label{tab:quantitative_comparison_simulation}
\end{table}
All methods besides SRAD and median show a similar performance with respect to MSE and MAD, where our method yields the best result.
In combination with the clearly varying image appearance of the different methods, this emphasizes the trade-offs between the methods compared here. 


\begin{figure}[htb]
     \centering
     \begin{subfigure}[b]{\columnwidth}
         \centering
         \includegraphics[width=0.7\textwidth]{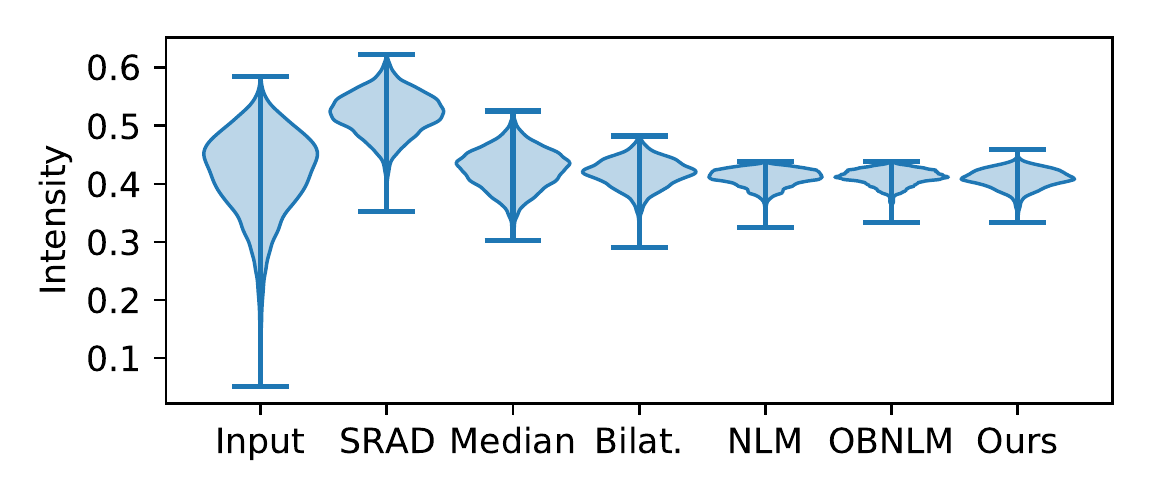}
         \caption{Background}
         \label{fig:quantitative_intensity_violin_experimental-a}
     \end{subfigure}
     \par
     \begin{subfigure}[b]{\columnwidth}
         \centering
         \includegraphics[width=0.7\textwidth]{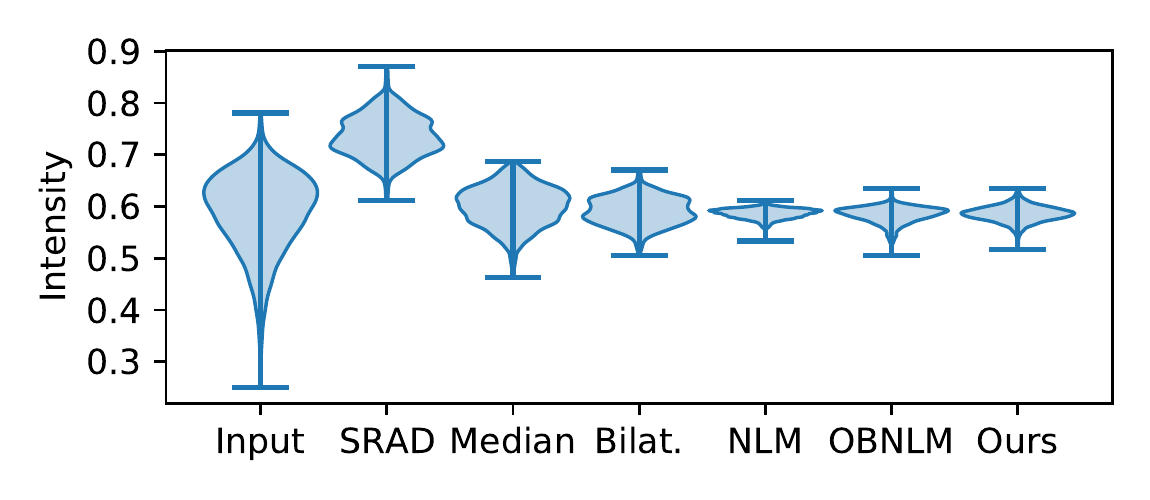}
         \caption{Hyperechoic Inclusion}
         \label{fig:quantitative_intensity_violin_experimental-b}
     \end{subfigure}
        \caption{Image intensity distribution within homogeneous background region (a) and hyperechoic inclusion (b) of the experimental images.}
        \label{fig:quantitative_intensity_violin_experimental}
\end{figure}
\begin{figure}[htb]
    \centering
    \includegraphics[width=0.7\columnwidth]{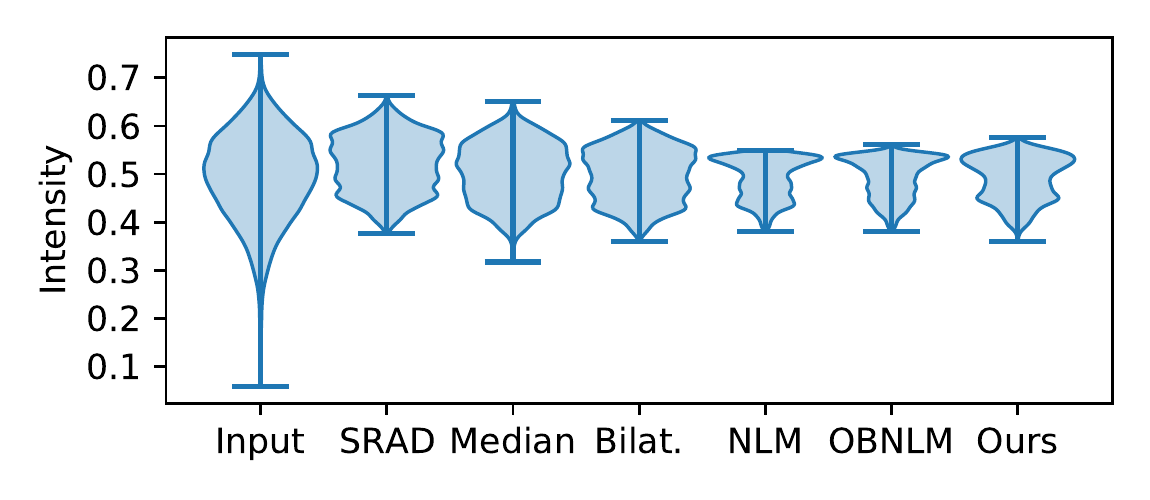}
    \caption{Image intensity distribution within homogeneous thyroid region of the in-vivo images.}
    \label{fig:quantitative_intensity_violin_invivo}
\end{figure}

\emph{Phantom and in-vivo data.}
In the experimental and in-vivo images it is not possible to observe different speckle instantiations and thus no average image can be computed.
Consequently, we focus on homogeneous image regions, namely the hyperechoic inclusion and the background for the phantom dataset (see \autoref{fig:qualitative_comparison_experimental-a}) as well as the thyroid in the in-vivo dataset (see \autoref{fig:qualitative_comparison_invivo-a}).
Within those regions of assumed homogeneous scatterer distribution the ideal speckle removal filter would result in a flat response.
\autoref{fig:quantitative_intensity_violin_experimental} shows violin plots of the intensity distribution in the marked regions of the phantom images, while \autoref{fig:quantitative_intensity_violin_invivo} shows those of the in-vivo image regions.
The results of NLM, OBNLM and our method on the experimental image show very similar distributions; SRAD, median and bilateral-filter however are less tightly packed.
This generally is matched by the evaluation for the in-vivo image data, but the distribution of our method is less pronounced than NLM and OBNLM.

In \autoref{tab:quantitative_comparison_experimental_invivo}, we show the image intensity standard deviations within the homogeneous image regions.
As expected from the violin plots, SRAD, median, and the bilateral-filter exhibit larger standard deviations in the three regions than NLM, OBNLM and our method.
While our method does not outperform NLM and OBNLM---which share best performance within the three regions and were tuned on the thyroid image---it shows how well our method translates to real ultrasound data, despite not having been trained with real images.
%
\begin{table}[htb]
\sisetup{detect-weight}
\centering
\begin{tabular}{l|ll|l}
\hline
\multicolumn{1}{l|}{Region} & \multicolumn{2}{l|}{Experimental  ($\times 10^{-2}$)}                       & \multicolumn{1}{l}{In-Vivo  ($\times 10^{-2}$)} \\ \hline
\multicolumn{1}{l|}{Method} & \multicolumn{1}{l}{Background} & \multicolumn{1}{l|}{Inclusion} & \multicolumn{1}{l}{Thyroid} \\ \hline
Input                       & 7.10               & 7.58           & 8.96                    \\
SRAD                        & 3.55               & 3.96           & 5.89                    \\
Median                      & 3.37               & 3.83           & 6.05                    \\
Bilateral                   & 2.39               & 2.69           & 5.42                    \\
NLM                         & 1.49               & \bfseries 0.96 & 4.21                    \\
OBNLM                       & \bfseries 1.32     & 1.60           & \bfseries 4.20          \\
Ours                        & 1.61               & 1.51           & 4.54                    \\ \hline
\end{tabular}
\caption{Quantitative evaluation on the experimental and in-vivo images and shown in \autoref{fig:qualitative_comparison_experimental-a} and \autoref{fig:qualitative_comparison_invivo-a} respectively. The table shows the image intensity standard deviation in homogeneous regions in the background and foreground of the experimental images, and within the thyroid in the in-vivo case.
Lowest standard deviation per region highlighted in bold without implying significance.}
\label{tab:quantitative_comparison_experimental_invivo}
\end{table}

\section{Discussion}
The qualitative and quantitative evaluations show that Speckle2Speckle works well in the simulated images, performing on-par with or even outperforming NLM and OBNLM, as can be seen in \autoref{fig:qualitative_comparison_simulation} and \autoref{tab:quantitative_comparison_simulation}.
It also performs well in real acquisitions, although no real images have been used for training. 
There are shortcomings however, especially regarding the overall smoothness as in the background area and the reproduction of the point scatterers in \autoref{fig:qualitative_comparison_experimental}.
Yet, it seems the highlighting of interfaces translates well to real images, with the exception of the linear structures in the muscle layer above the carotid artery in \autoref{fig:qualitative_comparison_invivo}.

    
The execution time of our method is significantly lower than for NLM and OBNLM.
While the used NLM and OBNLM implementations (CPU only) take in the range of minutes to execute, our method can be performed within tens of milliseconds on a recent GPU, making it suitable for real-time use.

\section{Conclusion}
We have shown with Speckle2Speckle a technique for training despeckling filters purely on simulated data as well as the generation of the simulated data.
The concept is intriguing, since it allows for the generation of multiple speckle realizations from the same underlying geometry.
The learned filter translates well to real acquisitions, performing similar to NLM and OBNLM, but only requiring a fraction of the execution time.

Since no clean training data is required, the number of simulations is drastically reduced compared to approaches that would use average images were used as target images.
It also enables creation of matching maps with high-level info on the structures present in the image that can be used to guide the outcome of the method during training---as we have done with interfaces in this case.

Including single point scatterers and/or strings in the simulated images could improve the reproduction of those structures.
They could be included in the interface map used for sharpening or placed in a separate map all-together, only depending on the desired appearance.
This aspect of using additional geometric maps to tune the result for particular locations could possibly be exploited beyond speckle removal, for example for clutter filtering in blood vessels within the same filter pass.


\acks{We acknowledge the important conversations with our colleagues Maximilian Baust, Maria Tirindelli, and Stefan Wörz.}

%
\ethics{The work follows appropriate ethical standards in conducting research and writing the manuscript, following all applicable laws and regulations regarding treatment of animals or human subjects.}

\coi{We declare we do not have conflicts of interest.}

\bibliography{references}

\begin{thebibliography}{15}
\providecommand{\natexlab}[1]{#1}
\providecommand{\url}[1]{\texttt{#1}}
\expandafter\ifx\csname urlstyle\endcsname\relax
  \providecommand{\doi}[1]{doi: #1}\else
  \providecommand{\doi}{doi: \begingroup \urlstyle{rm}\Url}\fi

\bibitem[Buades et~al.(2005)Buades, Coll, and Morel]{NLM_buades2005non}
Antoni Buades, Bartomeu Coll, and J-M Morel.
\newblock A non-local algorithm for image denoising.
\newblock In \emph{2005 IEEE Computer Society Conference on Computer Vision and
  Pattern Recognition (CVPR'05)}, volume~2, pages 60--65. IEEE, 2005.

\bibitem[Cammarasana et~al.(2021)Cammarasana, Nicolardi, and
  Patan{\`e}]{ESAOTE_cammarasana2021universal}
Simone Cammarasana, Paolo Nicolardi, and Giuseppe Patan{\`e}.
\newblock A universal deep learning framework for real-time denoising of
  ultrasound images.
\newblock \emph{arXiv preprint arXiv:2101.09122}, 2021.

\bibitem[Coup{\'e}(2011)]{OBNLM_code}
Pierrick Coup{\'e}.
\newblock Optimized bayesian non-local means filter, matlab package.
\newblock
  \url{https://web.archive.org/web/20201015011836/https://sites.google.com/site/pierrickcoupe/softwares/denoising-for-medical-imaging/speckle-reduction/obnlm-package},
  2011.
\newblock [Online; accessed 7-December-2021].

\bibitem[Coup{\'e} et~al.(2009)Coup{\'e}, Hellier, Kervrann, and
  Barillot]{OBNLM_coupe2009nonlocal}
Pierrick Coup{\'e}, Pierre Hellier, Charles Kervrann, and Christian Barillot.
\newblock Nonlocal means-based speckle filtering for ultrasound images.
\newblock \emph{IEEE Transactions on Image Processing}, 18\penalty0
  (10):\penalty0 2221--2229, 2009.

\bibitem[G{\"o}bl et~al.(2018)G{\"o}bl, Navab, and
  Hennersperger]{Goebl2018supra}
R{\"u}diger G{\"o}bl, Nassir Navab, and Christoph Hennersperger.
\newblock Supra: open-source software-defined ultrasound processing for
  real-time applications.
\newblock \emph{International Journal of Computer Assisted Radiology and
  Surgery}, Mar 2018.
\newblock ISSN 1861-6429.
\newblock \doi{10.1007/s11548-018-1750-6}.

\bibitem[Jensen(1996)]{jensen1996}
J{\o}rgen~Arendt Jensen.
\newblock Field: A program for simulating ultrasound systems.
\newblock In \emph{Proc. of the 6th Nordic Conf. on Human-Computer
  Interaction}, Medical \& Biological Engineering \& Computing, pages 351--353.
  Springer, 1996.

\bibitem[Jensen and Svendsen(1992)]{jensen1992}
J{\o}rgen~Arendt Jensen and Niels~Bruun Svendsen.
\newblock Calculation of pressure fields from arbitrarily shaped, apodized, and
  excited ultrasound transducers.
\newblock \emph{IEEE Transactions on Ultrasonics, Ferroelectrics, and Frequency
  Control}, 39\penalty0 (2):\penalty0 262--267, 1992.

\bibitem[Krull et~al.(2019)Krull, Buchholz, and
  Jug]{NOISE2VOID_krull2019noise2void}
Alexander Krull, Tim-Oliver Buchholz, and Florian Jug.
\newblock Noise2void-learning denoising from single noisy images.
\newblock In \emph{Proceedings of the IEEE/CVF Conference on Computer Vision
  and Pattern Recognition}, pages 2129--2137, 2019.

\bibitem[Lehtinen et~al.(2018)Lehtinen, Munkberg, Hasselgren, Laine, Karras,
  Aittala, and Aila]{Lehtinen2018}
Jaakko Lehtinen, Jacob Munkberg, Jon Hasselgren, Samuli Laine, Tero Karras,
  Miika Aittala, and Timo Aila.
\newblock {N}oise2{N}oise: Learning image restoration without clean data.
\newblock In Jennifer Dy and Andreas Krause, editors, \emph{Proceedings of the
  35th International Conference on Machine Learning}, volume~80 of
  \emph{Proceedings of Machine Learning Research}, pages 2965--2974,
  Stockholmsmässan, Stockholm Sweden, 10--15 Jul 2018. PMLR.

\bibitem[{Mohd Sagheer} and George(2020)]{DENOISING_REVIEW_sagheer2020review}
Sameera~V. {Mohd Sagheer} and Sudhish~N. George.
\newblock A review on medical image denoising algorithms.
\newblock \emph{Biomedical Signal Processing and Control}, 61:\penalty0 102036,
  2020.
\newblock ISSN 1746-8094.
\newblock \doi{https://doi.org/10.1016/j.bspc.2020.102036}.

\bibitem[Ronneberger et~al.(2015)Ronneberger, Fischer, and
  Brox]{UNET_ronneberger2015u}
Olaf Ronneberger, Philipp Fischer, and Thomas Brox.
\newblock U-net: Convolutional networks for biomedical image segmentation.
\newblock In \emph{International Conference on Medical Image Computing and
  Computer-Assisted Intervention (MICCAI'15)}, pages 234--241. Springer, 2015.

\bibitem[Tomasi and Manduchi(1998)]{BILAT_tomasi1998bilateral}
Carlo Tomasi and Roberto Manduchi.
\newblock Bilateral filtering for gray and color images.
\newblock In \emph{Sixth International Conference on Computer Vision (IEEE Cat.
  No. 98CH36271)}, pages 839--846. IEEE, 1998.

\bibitem[{Virginia Image and Video Analysis, School of Engineering and Applied
  Science, University of Virginia}(2006)]{SRAD_code}
{Virginia Image and Video Analysis, School of Engineering and Applied Science,
  University of Virginia}.
\newblock {MATLAB} and {C} implementation of speckle reducing anisotropic
  diffusion ({SRAD}).
\newblock
  \url{https://web.archive.org/web/20171213062541/http://viva-lab.ece.virginia.edu/downloads.html},
  2006.
\newblock [Online; accessed 7-December-2021].

\bibitem[Yin et~al.(2020)Yin, Gu, Zhang, Gu, Nie, Feng, Ma, and
  Yuan]{OPTICAL_yin2020speckle}
Da~Yin, Zhongzheng Gu, Yanran Zhang, Fengyan Gu, Shouping Nie, Shaotong Feng,
  Jun Ma, and Caojin Yuan.
\newblock Speckle noise reduction in coherent imaging based on deep learning
  without clean data.
\newblock \emph{Optics and Lasers in Engineering}, 133:\penalty0 106151, 2020.

\bibitem[Yu and Acton(2002)]{SRAD_yu2002speckle}
Yongjian Yu and Scott~T Acton.
\newblock Speckle reducing anisotropic diffusion.
\newblock \emph{IEEE Transactions on image processing}, 11\penalty0
  (11):\penalty0 1260--1270, 2002.

\end{thebibliography}

\end{document}